\documentclass[%
 reprint, %draft,
superscriptaddress,
 amsmath,amssymb,
 aps,
 pra,
]{revtex4-1}

\usepackage{graphicx}% Include figure files
\usepackage{dcolumn}% Align table columns on decimal point
\usepackage{bm}
\usepackage[dvipsnames]{xcolor}
\usepackage{hyperref}
\usepackage{upgreek}

\hypersetup{colorlinks=true,citecolor={OliveGreen},linkcolor={Maroon}}

\begin{document}

\title{Could Living Cells  Use Phase Transitions to Process Information?}

\author{Arvind Murugan}
\affiliation{%
 Department of Physics, University of Chicago, Chicago, IL, USA
}%

\author{David Zwicker}
\affiliation{%
 Max Planck Institute for Dynamics and Self-Organization, G\"{o}ttingen, Germany
}%

\author{Charlotta Lorenz}
\affiliation{Department of Physics, Department of Materials Science And Engineering, Cornell University, Ithaca, NY, USA}%

\author{Eric R. Dufresne}
\affiliation{Department of Physics, Department of Materials Science And Engineering, Cornell University, Ithaca, NY, USA}%

\begin{abstract}
To maintain homeostasis, 
    living cells process information with networks of interacting molecules.
Traditional models for cellular information processing have focused on networks of chemical reactions between molecules.
Here, we describe how networks of physical interactions could contribute to the processing of information inside cells. 
In particular, we focus on the impact of biomolecular condensation, a  structural phase transition found in cells. 
Biomolecular condensation has recently been implicated in diverse cellular processes.
Some of these are essentially computational, including classification and control tasks.
We place these findings in the broader context of physical computing, an emerging framework for  describing how the native dynamics of nonlinear physical systems can be leveraged to perform complex computations.
The synthesis of these ideas raises questions about expressivity (the range of problems that cellular phase transitions might be able to solve) and learning (how these systems could adapt and evolve to solve different problems).
This emerging area of research presents diverse opportunities across molecular biophysics, soft matter, and physical computing.
\end{abstract}

\maketitle
% \tableofcontents

\section{Introduction}

\begin{figure}[t]
    \centering
        \includegraphics[width=1.0\columnwidth]{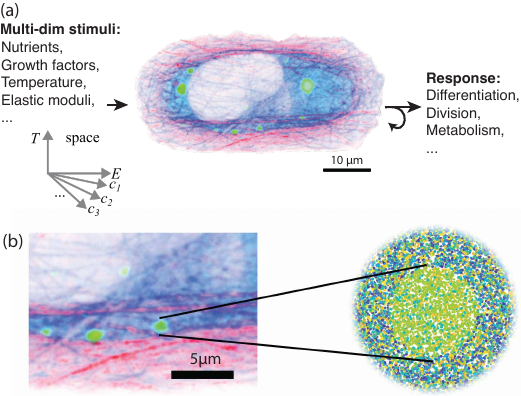}
    \caption{\emph{Living cells compute, potentially using biomolecular condensates.} 
    (a) Cells integrate diverse chemical and physical stimuli to determine their biological response. The shown U2OS cell is labelled for actin (pink), tubulin (blue), and the stress granule component G3BP1 (green)~\cite{boddeker2022non}.
    (b) Detail of panel (a) showing stress granules in green (left), which are fluid-like biomolecular condensates found in the cytoplasm that enrich many specific proteins and nucleic acids (schematic picture on the right).
    }
    \label{fig:reg}
\end{figure}

Living systems are far from thermodynamic equilibrium.
They achieve a healthy steady-state (\emph{i.e.} \emph{homeostasis}) 
across length-scales through energy-consuming closed-loop control.
Even an individual cell can sense and compute to mount appropriate responses to  external stimuli or internal changes of state, shown schematically in Fig.~\ref{fig:reg}a. Environmental cues could include sources of nutrients or stress \cite{desvergne2006transcriptional,wu2020regulatory,alfieri2007hyperosmotic}.
Internal signals can originate from many sources, including
stochastic expression of genes \cite{raser2005noise} or the  depletion of metabolites \cite{parry_bacterial_2014}. 

From an information perspective, homeostasis is a \emph{control} problem.
In some cases, cells respond to relatively simple `one-dimensionsal' stimuli such as the concentration of a single species.  For example, \emph{E. coli} and neutrophils move themselves up a concentration gradient in a process called chemotaxis \cite{weiner_regulation_2002,mattingly_escherichia_2021}.
More generally, cells process a vast array of physical (\emph{e.g.} matrix stiffness) and chemical signals (\emph{e.g.} soluble and matrix-bound growth factors). 
Processing of these high-dimensional stimuli can include \emph{classification}, the assignment of a label to a broad range of input values to summarize the relevant pattern in the signal. 
Classification allows cells to mount distinct responses to complex stimuli, such as advancement in the cell cycle  (growing, dividing, or going dormant), switching of mechanical state (free-floating, adhering, or crawling),  and differentiation into a specific cell type \cite{dongre2019new,bohnsack2004nutrient,scadden2006stem, engler2006matrix}. 
Indeed, classification has been recognized as an essential step in stem cell differentiation \cite{lang2014epigenetic,smart2023emergent}. 
The patterns triggering particular differentiation pathways require discrimination between many ligand combinations \cite{antebi2017combinatorial,su2022ligand}, 
and temporal patterns \cite{hansen2015limits,purvis2013encoding}.

Mechanisms for information processing are woven into the molecular fabric of life. 
At this scale, information processing  is often thought to be realized by complex networks of chemical reactions.  
Such \emph{regulatory pathways}  respond on short timescales through the interactions of proteins in the cytoplasm \cite{papin2005reconstruction}, and at longer timescales by coupling to gene expression in the nucleus \cite{lee2002transcriptional}.
While  some functional sub-networks can be neatly dissected from the whole \cite{babu2004structure}, this is usually not possible, as a typical node (\emph{i.e.} molecule)  can interact with many partners and contribute to many different functions \cite{huelsken2002wnt,clapham2007calcium}. 

In this Perspective, we discuss how two emerging topics in  physics -- biomolecular condensates  and physical computing -- could complement each other and provide fresh insights and new tools for information processing by molecular networks, particularly those inside living cells. 
After providing some concrete biological motivations, we describe how the basic physics of condensates could automatically achieve certain computational tasks, like classification and  control.
We then generalize these behaviors using essential ideas of physical computing and highlight open research questions.

\section{Computing by Biomolecular Condensates}

After briefly describing biomolecular condensates, a recently appreciated class of structures in cells, we describe how the physics of condensation can simply solve classification and control problems, and place these observations into the framework of physical computing.

\subsection{What are Biomolecular Condensates?}

Biological cells comprise thousands of different molecular components, which are organized into many spatial compartments. 
While some compartments are enclosed by lipid membranes (like the nucleus), \emph{biomolecular condensates}  are  domains enriched in specific proteins and nucleic acids that form spontaneously through cohesive interactions of their constituents.

Typical biomolecular condensates include many distinct components.
Stress granules, shown in Fig.~\ref{fig:reg}b, are thought to have hundreds of components~\cite{Youn2019}.
These components are enriched selectively based on physical interactions, which are dictated by specific sequences and chemical modifications (like oligomerization and post-translational modifications)~\cite{Holehouse2023, Choi2020a}.
While there has been  no systematic  census of condensates, merging the results of independent studies suggests that there are at least tens of distinct coexisting condensates~\cite{Rostam2023}. 
The compositional variation across different types of biomolecular condensates yields comparments with distinct physical and chemical properties. 
Chemically, they can localize biochemistry through enrichment or depletion of enzymes \cite{banani2017biomolecular} and diverse small molecules, including ions and metabolites \cite{ambadi2024small, posey2024biomolecular}. 
Further, macromolecules could adopt novel conformations in condensates~\cite{farag2022condensates, galvanetto_extreme_2023, emmanouilidis_nmr_2021} that could impact their reactivity. 
Physically, condensates have diverse rheological properties (\emph{i.e.} viscosity and elasticity), interfacial tensions, and electrostatic potential \cite{brangwynne_active_2011,Jawerth2020, ijavi_surface_2021,alshareedah2021programmable,dai_biomolecular_2024}. 

\begin{figure}
    \centering
    \includegraphics[width=\columnwidth]{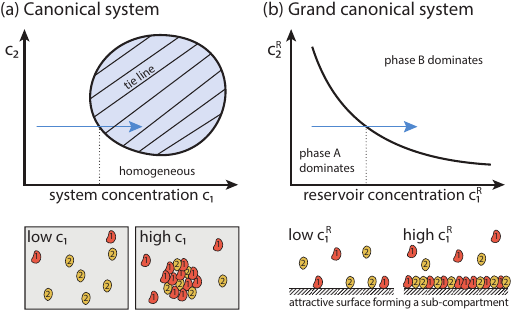}
    \caption{\emph{Phase diagrams of biomolecular condensates.}
    (a) Phase behavior of a closed system (canonical ensemble) as a function of the concentrations $c_1$ and $c_2$ of two components.
    In this example of associative phase separation (light blue area), domains rich in both components form as $c_1$ is increased to cross the phase boundary (blue arrow).
    (b) Phase behavior of a sub-compartment coupled to a reservoir (grand canonical ensemble) as a function of the reservoir concentrations $c^\mathrm{R}_1$ and $c^\mathrm{R}_2$.
    The composition in the sub-compartment undergoes a discontinuous jump when $c^\mathrm{R}_1$ is increased to cross the phase boundary (blue arrow).
    }
    \label{fig:phase_diagrams}
\end{figure}

Condensation of these biological molecules can be conceptualized as \emph{phase separation}, where  a homogeneous mixture sorts itself into at least two different macroscopic domains.
For systems near equilibrium, phase separation is understood as a competition of translational entropy favoring a homogenous mixed state and cohesive interactions favoring segregation into different phases.
The equilibrium configuration is described by phase diagrams, which show the preferred state of the system as a function of external parameters (Fig.~\ref{fig:phase_diagrams}).

When we consider an entire cell on time scales where production and degradation are negligible, the total amount of material in the system is conserved and we work in the canonical ensemble. In that case, large sections of the phase diagram will correspond to coexistence of multiple phases, which can differ in concentration by orders of magnitude (Fig.~\ref{fig:phase_diagrams}a).
Coexistence  typically manifests as droplets enriched in a suite of molecules that are depleted in the surrounded `dilute' phase.
For example,  stress granules (Fig.~\ref{fig:reg}b),  are enriched with the protein G3BP1 among many other biomolecules.
Stress granules can form spontaneously when the concentration of G3BP1 crosses a threshold (blue arrow in Fig.~\ref{fig:phase_diagrams}a).
Consequently, increasing the concentration of one component ($c_1$ in the figure) can induce droplets that locally increase the concentrations of many components (lower right panel in Fig.~\ref{fig:phase_diagrams}a).

Many condensates in cells form in small compartments, \emph{e.g.}, on membranes or at specific anchor sites~\cite{Zeng2016-hl, Shelby2023-tn}.
If components are exchanged with the surroundings, 
the surroundings can be conceptualized as a reservoir maintaining fixed concentrations, which corresponds to the grand canonical ensemble (Fig.~\ref{fig:phase_diagrams}b).
In the associated phase diagram, regions of coexisting phases collapse into phase boundaries -- off these boundaries, the system is either in one phase or another.
For example, only a few molecules will associate with the compartment (low $c_1$ and $c_2$) if the concentration $c_1^\mathrm{R}$ in the surrounding bulk is low (lower left panel in Fig.~\ref{fig:phase_diagrams}b).
When $c_1^\mathrm{R}$ is increased across the phase boundary (blue arrow in Fig.~\ref{fig:phase_diagrams}b), both concentrations $c_1$ and $c_2$ suddenly increase across the entire compartment (lower right panel).
This occurs when affinity toward the compartment and cohesive interactions of the components combine to  overcome the entropic penalty of confinement. 

In the following sections, we will describe how multi-component phase separation could achieve complex computational tasks.

\begin{figure}
    \centering
    \includegraphics[width=\columnwidth]{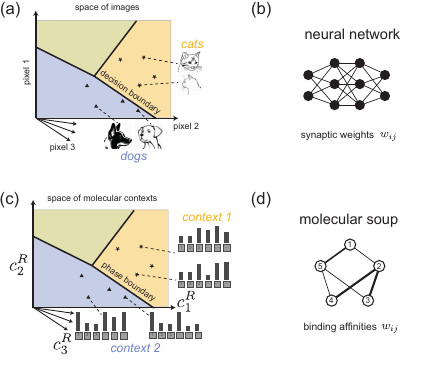}
    \caption{\emph{Phase boundaries in multicomponent systems as decision boundaries} (a) 
    A classification problem is solved by the right set of decision boundaries in the space of inputs (\emph{e.g.}, images). Each point in this diagram represents an image; the color denotes the output of classification (\emph{e.g.}, discrete classes such as `cat' or `dog'). (b) Decision boundaries can be placed in the desired locations by tuning the parameters $w_{ij}$ of a neural network. (c) In analogy, we interpret phase boundaries of multi-component phase separation as decision boundaries. Each point in this diagram represents a high-dimensional molecular input, \emph{i.e.},  concentrations $c_i$ of several molecular species. Molecular inputs on different sides of a phase boundary trigger the formation of distinct phases (outputs). (d) Phase boundaries can be placed as desired by tuning interaction parameters $w_{ij}$ of molecular mixtures. 
    }
    \label{fig:boundaries}
\end{figure}

\subsection{Condensation and Classification}

Classification  is a challenging computational task, where 
 a high-dimensional input, $\vec x$,  such as an image, is assigned  to a discrete category, $y$, known as a `class'.
A classic example is the discrimination of cats and dogs from images.
A single image corresponds to a  point $\vec x$ in a  high-dimensional space of possible images,  shown schematically in Fig.~\ref{fig:boundaries}a.
Since nearby $\vec x$ typically correspond to the same class, the mapping $\vec x \mapsto y$ can be visualized by decision boundaries, defined as regions where $y$ changes for small variations in $\vec x$.

Artificial neural networks have emerged as a powerful means to classify high-dimensional data~\cite{LeCun2015-lq}. They are represented by a set of synaptic weights $w_{ij}$ that describe interactions between the `neurons' $i$ and $j$ in the network (Fig.~\ref{fig:boundaries}b). 
To achieve a particular classification task, the weights $w_{ij}$ are adjusted in a process called `training'. Typically, artificial neural networks are implemented in software on generalized computing platforms.  

An analogous picture holds for multi-component phase separation in the grand canonical ensemble,  represented by the schematic phase diagram in Fig.~\ref{fig:boundaries}c. 
A point $\vec x$ in a multi-dimensional phase diagram represents reservoir concentrations of the relevant `input' species.
Crossing a phase boundary corresponds to a discontinuous change in structure and properties of the mixture, 
\emph{i.e.}, the formation of distinct phase $y$.

Comparing Figs.~\ref{fig:boundaries}a and~\ref{fig:boundaries}c, we re-interpret phase transitions in a multi-component system as  classification of high-dimensional chemical compositions. 
Intriguingly,  mean-field models of multi-component phase separation~\cite{zwicker_evolved_2022} are analogous to artificial neural networks.  
They are built upon a network representation of molecular interactions,  where  a set of  parameters $w_{ij}$  describe the physical coupling of different molecular species, shown in Fig.~\ref{fig:boundaries}d.
Just as $w_{ij}$ adjusts decision boundaries in an artificial neural network,   $w_{ij}$ moves phase boundaries in a multi-component system.

\subsection{Condensation and Control} 

Closed-loop control is an essential element of robust functional systems.
In a typical control system, one uses feedback regulation to maintain some aspect of the system --- the controlled variable $z$ --- at a set point 
in the face of fluctuations in a high-dimensional external input $\vec x$ \cite{aastrom2021feedback}.
For example, a chemostat can maintain a fixed pH within a reactor, even as non-stationary chemical reactions consume and produce pH-active species. 
An engineered control system has subsystems for sensing, computation, and actuation. 
A sensing system transduces the system state into an abstract signal. 
A computational system processes signals from sensors, calculates the desired response, and communicates it to actuators.  
In the chemostat example, an pH sensor transduces the pH into a voltage, a microprocessor processes the pH history, and communicates with an injector to introduce chemical species that offset the reaction's effect on the pH.  

Phase separation can greatly simplify control into a single physical process, without separate entities to sense, compute, and actuate.
In a widely discussed example, 
phase coexistence can suppress concentration  fluctuations \cite{banani2017biomolecular}. 
Consider a molecule whose total concentration fluctuates (due to factors like cell division or stochastic gene expression), which serves as an external input or disturbance $x$. 
Phase coexistence in the canonical ensemble suppresses these fluctuations by maintaining stable concentrations $z$ within each of the coexisting phases. 
To absorb changes to the total concentration, the system adjusts the fraction of volume occupied by either phase. 
This  mechanism is well-established in simple one- and two-component systems and has been observed in some strongly self-segregating cellular components \cite{klosin2020phase}. However, it remains unclear whether such robust buffering through phase separation extends to condensates containing many components without requiring fine-tuning~\cite{deviri2021physical}.

\subsection{Examples from Living Cells}

Cells utilize biomolecular condensates to trigger discrete courses of action, \emph{e.g.}, to decide which genes to express.
Gene expression requires the assembly of many proteins (RNA Polymerase, mediator complex, and others) at a particular location on the chromatin to initiate transcription~\cite{Sabari2018-pj,Boija2018-gk}.
The assembly is triggered by the presence of regulating molecules (transcription factor, coactivators, and enhancers), which recognize a motif on the chromatin.
The formation of such transcriptional condensates is a classification, which converts the high-dimensional regulatory input into a discrete transcriptional
outcome.

The immune system provides another example for classification \emph{via} phase separation:
when T cell receptors are exposed to the right combination of input features on an antigen-presenting cell (\emph{e.g.}, antigen affinity, dwell time, co-stimulatory ligands, and local kinase activity), signaling proteins condense at the immunological synapse~\cite{Su2016-io,Shimobayashi2021-us}.
This drives full T cell activation, effectively classifying diverse ligand inputs into an all-or-none immune response.

Condensates also naturally help the cell solve control problems since the reorganization of molecules into condensates allows cells to  maintain homeostasis. Compared to other regulatory mechanisms such as transcription, translation, or protein degradation, condensation is more rapid and metabolically efficient. For example, most pre-stress mRNAs and many RNA-binding proteins (\emph{e.g.}, translation initiation factors, poly(A)-binding proteins) condense into stress granules~\cite{Yoo2019-xz,protter2016principles} in response to stress. This sequestration reduces bulk translation, freeing up ribosomes to prioritize newly transcribed messages critical for stress survival, while preserving older transcripts for later reuse~\cite{Glauninger2024}.
In this way, stress granules rapidly rewire translation without requiring new transcription.
A similar sequestration is observed for ribosomal subunits that cannot form a complete ribosome~\cite{Ali2023-zo}; such sequestration provides a more metabolically efficient mode of regulation than protein degradation.

Similarly, condensates allow for quick response time when unfolded proteins accumulate in the ER (the unfolded protein response)~\cite{Hetz2020-ao}. These unfolded proteins trigger a condensate composed of an enzyme (IRE1) and its substrate (specific mRNA to be spliced) that alleviates that stress.
Here, condensates allow for quick on and off response to stress by forming and dissolving the condensate that co-localizes enzyme and substrate. 

\subsection{Physical Computing and Condensates}

\begin{figure}
    \centering
    \includegraphics[width=.7 \columnwidth]{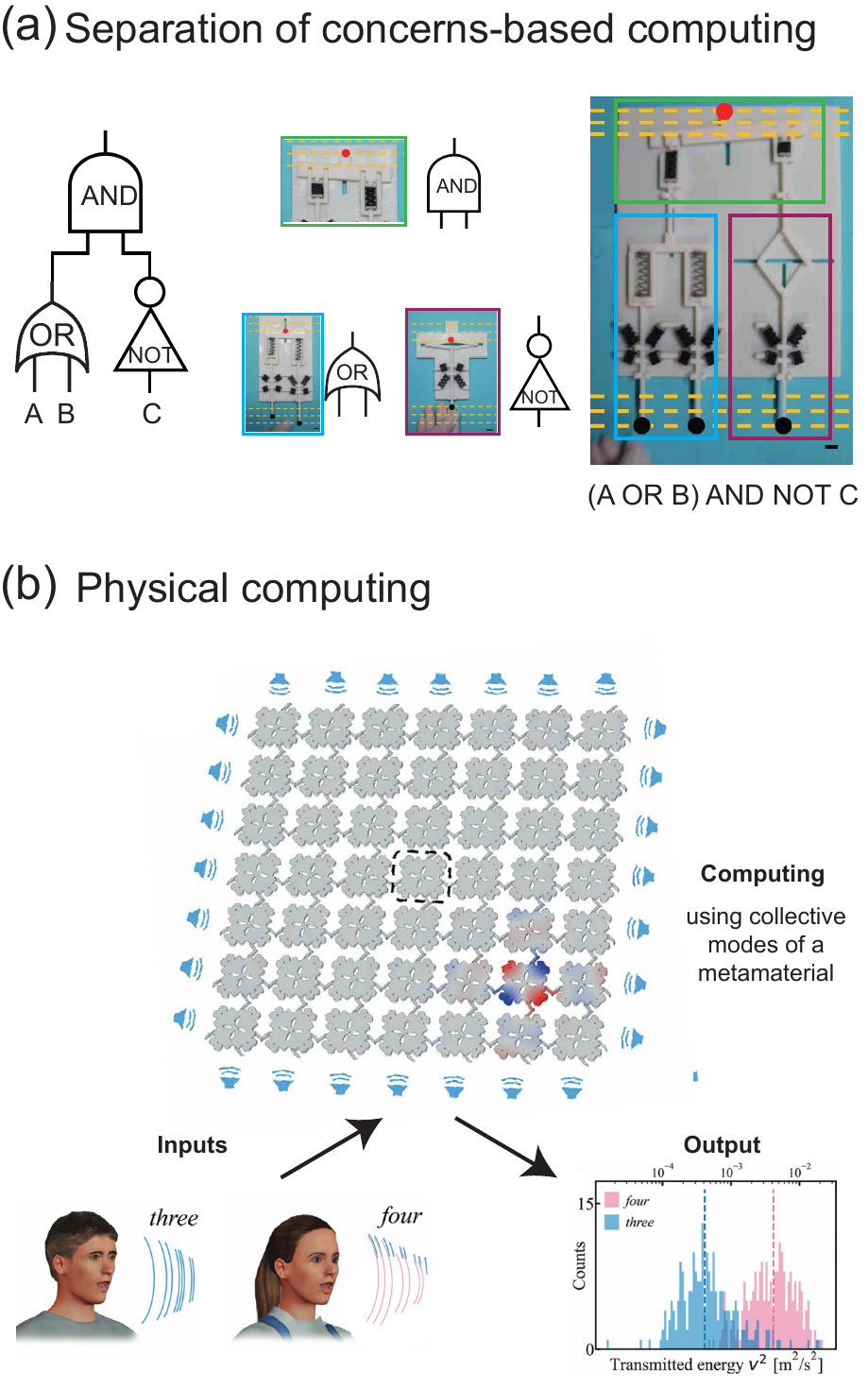}
    \caption{\emph{Contrasting Separation of Concerns and Physical Computing in Mechanical Systems}  (a) A complex logical function is broken down into a network of elementary logic gates.  Modular mechanical devices mimic logic gates and are assembled  to realize the desired input/output relationship \cite{Zhang2021-bk}. (b)~The structure of an acoustic resonator with many degrees of freedom is optimized to classify audio input, separating the words `three' and `four'. Modified from \cite{dubvcek2024sensor}.   
    }
    \label{fig:comp}
\end{figure}

Our discussion has demonstrated the potential for phase separation to achieve classification and control.
In this section, we place these observations in a broader context, and contrast two approaches to achieving computation in a physical system: (a) classic systems engineering based on `separation of concerns'~\cite{Dijkstra1982-jd,Hartwell1999-io,Endy2005-yb}, and (b) physical computing, inspired by `analog computing' ~\cite{Brooks1991-pu,ulmann2013analog, braitenberg1986vehicles, bull2021excitable,Laydevant2023-mn}. 

A `separation of concerns' approach emphasizes abstraction and general solutions that are not tied to 
the physics idiosyncratic to a particular system (Fig~\ref{fig:comp}a). Typical examples in biomolecular computation include systems and synthetic biology networks such as circuits of logic gates implemented with molecules. With this strategy, a control task might be split into subtasks of sensing, computation, and actuating a response; 
 a layer of receptors might sense environmental `input' signals, a well-mixed molecular network \emph{e.g.}, phosphorylation/de-phosphorylation network) might carry out the computation, and finally a downstream physical process (gene expression, self-assembly, condensation) is actuated as the output \cite{Bray1995-yf, Hartwell1999-io,Erbas-Cakmak2018-rm,Nandagopal2011-dc,Qian2011-jh,Buchler2003-xf}. This approach tends to produce modular solutions, with independent sub-systems that complete discrete sub-tasks.  This approach enables the familiar process of writing high-level computer code that can be implemented across a range of hardware platforms.  It can be applied, for example, to the creation of an artificial neural network for classification. 
 
An alternative physics-first approach -- physical computing -- does not separate software from hardware \cite{ulmann2013analog, Laydevant2023-mn,Brooks1991-pu}.  
Instead, it exploits the intrinsic physics of a specific system to solve a problem (Fig~\ref{fig:comp}b). 
Physical computing solutions have been developed for classification problems in the optical \cite{mcmahon2023physics}, electronic \cite{wright2022deep}, mechanical \cite{stern2020supervised}, and molecular domains \cite{Woods2019-ke,winfree1998algorithmic, evans2024pattern}.
For example, in the mechanical domain, the vibrational modes of a non-periodic phononic metamaterial can be trained to recognize  certain spoken words \cite{dubvcek2024sensor}, as shown in Fig~\ref{fig:comp}b.
In the molecular context, the kinetics of nucleation and growth in DNA self-assembly can classify patterns in the concentration of a many-component mixture \cite{evans2024pattern}.

Physical computing generally affords compelling advantages. Physical computing systems tend to be energy efficient, with the input signal often providing the energy needed for the computation \cite{wright2022deep, evans2024pattern, dubvcek2024sensor}. In addition, solutions tend to be compact, since the collective physics of a set of components can achieve sensing \cite{Yoo2019-xz}, many layers of computation, and response without dedicating separate modules to each role.  
For example, the physical computing approach to molecular pattern recognition in \cite{evans2024pattern} used phase boundaries as decision boundaries (as described in Sec II and Fig.~\ref{fig:boundaries}); environmental conditions that were to be classified into two different categories (with different molecular responses) were separated by placing a phase boundary (for crystallization) between those conditions. In this approach, the entire process of sensing, computation, and response is achieved by one physical process, namely the nucleation and growth of a specific phase in a specific part of the phase diagram.

The main downside of physical computing is that solutions cannot be devised in the abstract, independent of the physics of the system. This aspect is a limitation for general purpose computing~\cite{Laydevant2023-mn}, but in the context of biology, parsimony, compactness, lower energy costs, and robustness appear to give physical computing a decisive advantage.

\begin{figure*}[htbp]
    \centering
    \includegraphics[width=\textwidth]{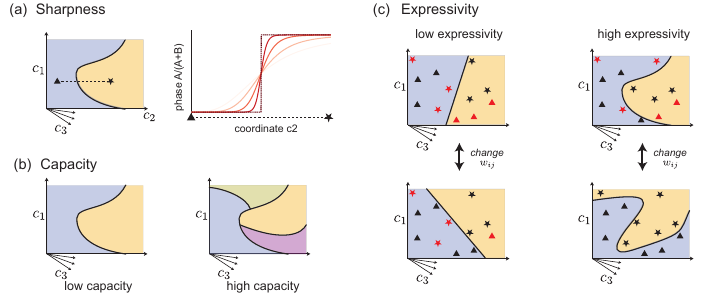}
    \caption{\emph{Quantifying computational capabilities of phase diagrams}  (a) Sharpness of phase boundaries can be quantified by the maximal slope of the response  as one crosses a phase boundary; low sharpness can limit the decision quality by producing mixed or ambiguous outcomes. 
    (b) Capacity, the number of distinct phases in a phase diagram, limits the number of distinct output categories.
    (c) Expressivity quantifies the ability to classify diverse data sets by tuning interaction parameters $w_{ij}$. 
    Stimuli corresponding to triangles and stars are meant to evoke two different phases (blue and yellow respectively) as responses. Black and red symbols denote stimuli that evoke correct and incorrect responses respectively.} 
    \label{fig:class}
\end{figure*}

\section{Classification power of condensates}

In this section, we discuss the potential for phase separating systems to classify complex high-dimensional inputs. After describing the classification problem in  more detail, we discuss  open questions in classification by  phase separation. 

\subsection{Quantifying computational abilities}

Just as physicists have metrics for physical properties (transport coefficients, elastic moduli), computer scientists have developed metrics for quantifying computational abilities of a system. 
Here, we describe metrics developed to quantify the ability of systems to solve diverse classification problems.

\textbf{Sharpness:} The most elementary property of a decision boundary relevant to information processing is its sharpness, \emph{e.g.}, characterized by the maximum slope of sigmoidal response curves in Fig.~\ref{fig:class}a.
A shallow slope implies a graded response, which can reduce computational precision by introducing ambiguity near the decision threshold. In contrast, sharp transitions can enable decisive, high-fidelity outcomes. Idealized phase transitions in the grand canonical ensemble are infinitely sharp.  In real molecular systems, sharpness can be limited by finite system size and kinetic effects (described below). 

\textbf{Capacity:} We define capacity as the number of potential outputs that a system can produce; it limits the number of distinct categories that high-dimensional inputs can be mapped onto. In a wide family of recurrent neural network models inspired by Hopfield's work~\cite{Hopfield1982-yy}, capacity quantifies the number of distinct stable attractor states that a system can support.
In these systems, capacity is known to scale with the number of neurons and depends on the architecture of the underlying network~\cite{Krotov2020-ka,Hertz1991-sg}. For phase separation in the grand canonical ensemble,  capacity can be quantified by the number of distinct phases in a phase diagram as reservoir concentrations are varied (\emph{e.g.}, like that shown in Fig.~\ref{fig:class}b). Such a capacity has been quantified for crystalline phases \cite{murugan2015multifarious}, polymer states \cite{Fink2001-zf}  and, most relevant to this perspective, for liquid phases \cite{Braz-Teixeira2024-jl, Jacobs2021-zn}. 
In the canonical ensemble, multiple phases can coexist.  In this case, capacity is greatly expanded, and can be quantified as the number of distinct combinations of coexisting phases.

\textbf{Expressivity:} Expressivity quantifies the range of classification tasks a system can perform or, more broadly, the system’s ability to compute outputs from inputs (\emph{e.g.}, approximating arbitrary functions~\cite{Hornik1989-by}). There are multiple measures of expressivity~\cite{Raghu2016-bw,Kearns2019-dn}, but they all capture the range of input-output relationships that a system can encode. For example, given an input space and an output space, a more expressive system can solve more complex classification tasks as system parameters $w_{ij}$ are varied; see Fig.~\ref{fig:class}c. 
One quantification of expressivity (Vapnik–Chervonenkis (VC) dimension
\cite{Kearns2019-dn})  
relies on determining the most complex classification problem a particular family of systems can solve (how many arbitrarily chosen points can it discriminate as belonging to one of two classes); more expressive systems can distinguish sets with more points. Expressivity correlates with the possibility of placing decision boundaries flexibly in the input space as parameters of the underlying system (\emph{e.g.}, neural network weights $w_{ij}$) are varied. For instance, deeper neural networks have higher expressivity under this metric~\cite{Raghu2016-bw}.
 
For phase separation, expressivity would be summarized by phase boundaries in a phase diagram, as shown in Fig.~\ref{fig:class}c. 
When physical parameters (\emph{e.g.}, binding affinities $w_{ij}$) are optimized, systems with higher expressivity can place their decision boundaries to accommodate more complex input patterns than less expressive systems, \emph{e.g.}, deploying different responses to stimuli that differ only through subtle higher order correlations; see Fig.~\ref{fig:class}c.

\subsection{Physical determinants of computational abilities}

Research over the years has shown how neural network architecture impacts computational abilities as quantified by these metrics; \emph{e.g.}, deeper networks are more  expressive.
However, we do not currently understand how the physics of multicomponent phase separation determines its computational abilities. Here, we highlight some key questions and preliminary work done towards this question and work that remains to be done.

\textbf{Hidden components as hidden neurons:}
Phase separation can involve molecular species that are neither `inputs' nor `output' molecules that output phases are enriched in. These  `hidden' species can mediate more complex phase behavior by interacting and co-condensing with the input and output molecules~\cite{stern2020supervised,evans2024pattern,Parres-Gold2023-pb,Chalk2024-em}. 
Consider a molecular system with $N$  species, where only 
$M$ species 
serve as inputs and outputs. Even if the full $N \times N$ interaction matrix $\omega_{ij}$ is fixed, 
the $M \times M$ apparent interaction matrix for the inputs and outputs depend on the concentrations of the $N-M$ hidden species. 
Thus, the concentrations of these hidden components can be tuned to expressively manipulate the positions of phase boundaries relative to the concentrations of `input' species. In this way, `hidden' molecules would play the role of `hidden' layers in deep neural networks that do not directly interface with input or output, but increase the expressivity of input-output maps \cite{Raghu2016-bw,Gunawardena2022-tg}. Since hidden molecules interact in a reciprocal manner (i.e., $\omega_{ij}=\omega_{ji}$), 
they are analogous to hidden neurons that increase expressivity in Boltzmann machines~\cite{Montufar2018-mh} rather than in feedforward networks with non-reciprocal interactions. 

\textbf{Nucleation boundaries for higher expressivity:} Phase boundaries, like that shown in Fig.~\ref{fig:class}, are usually understood to describe first-order equilibrium phase transitions. However, condensation outcomes are rarely  determined solely by equilibrium thermodynamics. Instead, they   could be dominated by nucleation. For example, certain concentration patterns of input species might preferentially promote the nucleation of one type of droplet over another, even if the latter is thermodynamically more stable. In this case, decision surfaces shown in Fig.~\ref{fig:class} are determined by the competitive nucleation of different phases. Despite not being thermodynamic boundaries, such kinetically-determined boundaries can nevertheless be sharp, especially when coupled to competition effects \cite{evans2024pattern,Parres-Gold2023-pb}.

Decision surfaces determined by kinetics are known to be more expressive than equilibrium phase boundaries for self-assembly \cite{zhong2017associative,evans2024pattern}.  For example, a region of the phase diagram with only one equilibrium phase can nevertheless show multiple kinetic phases. The stabilization of specific kinetic phases could be achieved through surfactant-like species that reduce nucleation barriers. 
Such computational benefits of nucleation in the context of liquid-liquid phase separation have yet to be systematically explored.

\textbf{Ripening for temporal information processing:}
Phase separation involves numerous transient phenomena. The finite timescales of these phenomena can be used for processing temporally coded information in input signals~\cite{purvis2013encoding}. One recent example exploits the kinetics of Ostwald ripening to deploy DNA damage repair programs in response to specific oscillatory waveforms in the upstream transcription factor p53~\cite{Heltberg2022}. A similar coarsening process also ensures a well-defined arrangement of droplets that determine where maternal and paternal genetic material is exchanged in germ cells of sexually reproducing organisms~\cite{Girard2023}. Finally, time delays inherent to nucleation of condensates of transmembrane proteins (LAT) are thought to be exploited for kinetic proofreading in T cell receptor (TCR) signaling pathway \cite{McAffee2022-kt, Huang2019-pi, White2025-wn}.   This time delay is used to prevent activation of T cells by low affinity ligands. We do not have a broader sense of the range of temporal information processing enabled by the many kinetic aspects of phase separation.

\textbf{Competition leads to sharper decisions:} 
Both equilibrium and non-equilibrium mechanisms can enhance decision sharpness in phase-separating systems near phase boundaries. At equilibrium, ensemble choice has a defining role. When the system is coupled to a large reservoir of components (approximating the grand canonical ensemble), sharp transitions in droplet composition and volume can emerge from small changes in reservoir concentrations. In contrast,  with fixed total concentrations (the canonical ensemble), equilibrium behavior yields more graded transitions, as droplets form gradually and grow continuously when control parameters (\emph{e.g.}, a component's concentration) cross a phase boundary. Real systems can lie somewhere in between, depending on the size of the reservoir relative to the condensing subsystem~\cite{Rossetto2024}.

Sharp decision-making is still possible in the canonical ensemble setting if kinetic effects are considered. For example, if multiple phases capable of coexistence share a limiting component, small differences in nucleation rates can be amplified. A phase with slightly faster nucleation can rapidly deplete the shared resource, suppressing droplet formation elsewhere and leading to a winner-take-all outcome. This kinetic competition can produce sharp, effectively irreversible decisions that go beyond what equilibrium thermodynamics predicts.

Such depletion-driven competition has been shown to enhance classification in molecular systems with low intrinsic cooperativity, such as BMP ligand gradients~\cite{antebi2017combinatorial,su2022ligand,Parres-Gold2023-pb} and other contexts \cite{Genot2012-ba, Kieffer2023-hr}. Similarly, recent work in multi-component self-assembly demonstrates that shared-component competition can sharpen phase boundaries beyond equilibrium limits~\cite{evans2024pattern}. While these effects have yet to be fully explored in the context of phase separation and computation, they suggest that even modest physical asymmetries can be harnessed to produce precise molecular decisions.

\textbf{Non-equilibrium dynamics for enhanced control:}
Computation by condensate networks could be further enhanced when biomolecular reactions are actively driven using a source of energy, such as ATP hydrolysis.
Such driven reactions feature prominently in known cellular regulatory networks, and their coupling to the cooperative physics of phase separation would extend their capabilities~\cite{Zwicker2022a}.
Driven reactions that affect droplet material can suppress nucleation and effectively shift phase boundaries~\cite{Ziethen2023,jambon2023phase}.
Cells could use this mechanism to suppress condensation until energy becomes scarce, which could trigger condensation in times of crisis~\cite{Wurtz2018a}.
Driven reactions can also control when and where droplets form, how large they become~\cite{Zwicker2015, Soeding2019, Kirschbaum2021}, and divide droplets when they are too large~\cite{Zwicker2017}.
Such delicate control over  morphology affects the physical interactions of droplets and thus also the integrated reaction network.

Chemical reactions coupled to phase separation can enable spatial control of pattern formation, as in non-ideal reaction networks~\cite{Carati1997, Aslyamov2023,Demarchi2023-ru,Menou2023}.
Driven reactions can maintain long-ranged compositional gradients, which could power information processing schemes.
For example, active \emph{in vitro} condensates have been shown to sense and `swim' toward each other, mediated through the chemical gradients they generate \cite{jambon2023phase}. 
Similarly, spatial gradients inherent to condensates can be exploited to power proofreading mechanisms that enhance classification accuracy~\cite{Galstyan2020-px}, without needing the molecular complexity required by traditional kinetic proofreading schemes \cite{Hopfield1974-wa}.

\subsection{Learning by Condensates}
A natural question raised by the idea of computation through phase transitions is how the relevant interaction parameters arise. 
While evolutionary processes can in principle tune biophysical interactions to generate functional phase diagrams, cellular adaptation often occurs on much faster timescales, requiring non-genetic mechanisms.

Recent work on `physical learning' suggests one possibility: materials can acquire new behaviors by experiencing examples during a `training' period \cite{Stern2023-fe,Gunawardena2022-tg}. However, this framework has faced a mechanistic challenge — how can physical interaction parameters, the analog of neural network weights, be updated through local rules without genetic change? One solution is post-translational modification of proteins, which can alter their effective interactions by changing charge, conformation, or binding affinities, thereby reprogramming phase behavior\cite{Hofweber2019-so,Wang2018-uv}. The concept of hidden components in multi-component phase separation, introduced earlier, provides another natural mechanism for training. In such systems, expression levels of hidden species modulate apparent interactions among inputs and outputs, even when the underlying interaction matrix is fixed \cite{evans2024pattern,Parres-Gold2023-pb,Chalk2024-em}. When components are densely and promiscuously connected, modifying expression levels alone can reconfigure phase behavior—enabling distinct computations in different cell types \cite{antebi2017combinatorial,su2022ligand,Parres-Gold2023-pb} or subcellular locations, without requiring mutations.
If expression levels are themselves influenced by the outcome of phase behavior, the system can, in principle, undergo Hebbian-like updates: for instance, phase separation may induce scaffolding protein expression \cite{Zhu2022-gm,evans2024pattern}, altering future interactions and behaviors. While explicit training rules remain to be determined, such architectures suggest a plausible route to local, non-genetic learning in molecular systems.

\section{Conclusion}
This perspective explores the potential of phase separation to process information in living cells.
While phase separation has already been implicated in essential regulatory processes~\cite{Alberti2019-rm,Yoo2019-xz,banani2017biomolecular},  this perspective presents a broader view inspired by artificial neural networks.
In lieu of the dominant systems biology perspective on signaling and gene regulation, which emphasizes modularity~\cite{Hartwell1999-io}, we advocate for a physical computing perspective, which emphasizes compact solutions that seamlessly integrate sensing, computation, and actuation.
Phase separation is well-suited to physical computing because it involves many species,  with a rich web of interactions. While physical computing solutions may not be interpretable or portable, they could be favored by  natural selection for their effectiveness and efficiency.

 The perspective requires much experimental and theoretical investigation to be fully realized. The main experimental challenges revolve around the development of tractable model systems, where input signals can be readily controlled, output signals can be reliably read out, and physical and chemical couplings can be measured and manipulated. 
We expect that \emph{in vitro} experiments with model systems will be necessary to reveal the essential capabilities and limitations of physical computing in biomolecular systems \cite{Baulin2025-qh,Gong2022-iu, Takinoue2023-cp, Udono2023-yu,Fabrini2023-ii,jambon2023phase}.
 While the basic physics of equilibrium phase separation  has long been understood, viewing multicomponent phase separation as a computation raises  new theoretical questions, which we have highlighted and  organized.
 This research program requires understanding how traditionally studied physical effects - kinetics and nucleation, active processes, and spatial patterning - impact computational abilities such as expressivity, memory capacity, sharpness of decision making. Solutions will integrate concepts from   computer science, information theory and physics, and will push the frontiers of active matter and biological physics.  

\section{Acknowledgements}
We thank Frank J\"{u}licher, Peter McMahon, Robert Style, Mason Rouches, Erik Winfree, Cameron Chalk, Krishna Shrinivas, Carla Fernandez-Rico, Jennifer Schwarz, William Jacobs, Gregory Carlson, and Richard Prum for their feedback on this manuscript.

\bibliography{ref,am_paperpile}
\end{document}